\documentclass[twocolumn,aps,floatfix,superscriptaddress]{revtex4}
\pdfoutput=1 
\usepackage{amsmath,amssymb,stackrel,eucal,graphicx,float,epstopdf}
\usepackage{mathtools}
\usepackage{subfiles}

\usepackage[colorlinks=true, urlcolor=blue, anchorcolor=blue, citecolor=blue,filecolor=blue,linkcolor=blue,menucolor=blue]{hyperref}

\begin{document}

\author{S. S. Budzinskiy}
\affiliation{Faculty of Computational Mathematics and Cybernetics, Lomonosov MSU, Moscow, Russia}
\affiliation{Institute of Numerical Mathematics RAS, Moscow, Russia}
\author{S. A. Matveev}
\affiliation{Faculty of Computational Mathematics and Cybernetics, Lomonosov MSU, Moscow, Russia}
\affiliation{Institute of Numerical Mathematics RAS, Moscow, Russia}
\author{P. L. Krapivsky}
\affiliation{Department of Physics, Boston University, Boston, MA 02215, USA}

\title{Hopf bifurcation in addition-shattering kinetics}

\begin{abstract}
In aggregation-fragmentation processes, a steady state is usually reached in the long time limit. This indicates the existence of a fixed point in the underlying system of ordinary differential equations. The next simplest possibility is an asymptotically periodic motion. Never-ending oscillations have not been rigorously established so far, although oscillations have been recently numerically detected in a few systems. For a class of addition-shattering processes, we provide convincing numerical evidence for never-ending oscillations in a certain region $\mathcal{U}$ of the parameter space. The processes which we investigate admit a fixed point that becomes unstable when parameters belong to $\mathcal{U}$ and never-ending oscillations effectively emerge through a Hopf bifurcation. 
\end{abstract}

\maketitle


Two complementary processes, aggregation and fragmentation, are widespread in Nature \cite{smol1917,BT45,Flory,agg-rev,skf,Icarus,cloud,book,Analytical}. Mathematically, a well-mixed system undergoing aggregation and fragmentation is described by equations 
\begin{eqnarray}
\label{AF:gen}
\frac{dc_k}{dt} &=& \frac{1}{2}\sum_{i+j=k} K_{ij}\,c_i\,c_j-
c_k\sum_{j\geq 1} K_{kj}\,c_j \nonumber\\
&+& \sum_{j\geq 1}F_{kj}c_{j+k} - \frac{1}{2}\,c_k\sum_{i+j=k} F_{ij}
\end{eqnarray}
Here $c_k(t)$ denotes the density of clusters composed of $k$ monomers, $K_{ij} = K_{ji} \geq 0$ is the rate of aggregation 
\begin{equation}
\label{K-def}
[i] \oplus [j]~\xrightarrow{\text{$K_{ij}$}} ~[i+j]
\end{equation}
and $F_{ij} = F_{ji} \geq 0$ is the rate of binary fragmentation 
\begin{equation}
\label{F-def}
[i+j]~  \xrightarrow{\text{$F_{ij}$}} ~[i] + [j]
\end{equation}

The system \eqref{AF:gen} of infinitely many non-linear ordinary differential equations (ODEs) is analytically intractable apart from a few special cases. The long-time behavior is easier to probe. If the mass distribution becomes stationary, one may guess the stationary distribution by equating the rate of the aggregation process $[i] \oplus [j]\to [i+j]$ to that of the reverse fragmentation process $[i+j] \to [i] + [j]$.  This detailed balance condition gives 
\begin{equation}
\label{balance}
 K_{ij}\,c_i\,c_j =  F_{ij}\,c_{i+j}
\end{equation}
The detailed balance condition determines the stationary distribution only in exceptional cases. Generically Eqs.~\eqref{balance} form an overdetermined system that does not possess a solution \cite{bk08}. 

More rich stationary states have been found in some systems amenable to analysis, e.g., in addition to a stationary distribution of finite clusters an infinite cluster comprising a finite fraction of mass of the entire systems is sometimes formed  (see \cite{kr,maj,ik,rm,jain}). Some aggregation-fragmentation processes are characterized by unlimited growth, namely the typical cluster mass diverges in the long time limit. Non-thermodynamic behaviors and non-equilibrium phase transitions have been also observed \cite{bk08}. These complicated behaviors reflect the peculiarities arising in infinitely many ODEs. 

The emergence of the stationary distribution is a more basic feature since it is merely the fixed point and fixed points often determine the long time behavior in systems of a few ODEs. For a single ODE, fixed points are crucial; for two coupled ODEs, the asymptotic behavior may be determined by a fixed point or a limit cycle. For more than two equations, chaos may emerge. Thus one would like to find never-ending oscillations and chaos in aggregation-fragmentation processes. 

Detecting a limit cycle in a system of two coupled ODEs is difficult \cite{Perko,Chris,Strogatz}. In Eqs.~\eqref{AF:gen}, the right-hand sides are quadratic polynomials. Finding limit cycles for the system
\begin{equation}
\label{PQ}
\frac{d x}{dt} = P(x,y), \quad \frac{d y}{dt} = Q(x,y)
\end{equation}
where $P$ and $Q$ are quadratic polynomials is the part of the Hilbert's 16th problem \cite{Hilbert}, see \cite{Ily02} for its fascinating history. More precisely, Hilbert asked (i) whether the number of limit cycles is finite for any polynomials $P$ and $Q$, and (ii) does it exist a universal upper bound $H(n)$ on the number of limit cycles depending only on the maximal degree $n=\text{max}[\text{deg}(P), \text{deg}(Q)]$.  The affirmative answer to the first question was established in \cite{Ily91,Ecalle92}. The answer to the second question is unknown, apart from the case of linear vector fields which have no limit cycles \cite{LC:def}, that is $H(1)=0$. For quadratic vector fields, systems with four limit cycles have been discovered \cite{H4_1,H4_2}. Thus $H(2) \geq 4$; so far, the possibility that $H(2)=\infty$ has not been ruled out. 


Persistent oscillations have been numerically observed in \cite{Colm12} for some open aggregating systems driven by input at small masses and sink at large masses. Oscillations could be caused by the drive, however. In closed systems, never-ending oscillations have been numerically detected in a class of processes with collision-controlled fragmentation where each fragmentation event leads to complete shattering of colliding clusters into monomers:
\begin{equation}
\label{S-def}
[i] \oplus [j]~ \xrightarrow{\text{$S_{ij}$}} ~\underbrace{[1]+\cdots+[1]}_{i+j}
\end{equation}
Since the binary collision can lead to aggregation or shattering, the reaction rates that differ only by an amplitude, $S_{ij} = \lambda K_{ij}$, have been explored \cite{pnas2015,AS17,AS18,Colm18,sab18,Newton19}. For the family of rates $K_{ij} = (i/j)^a + (j/i)^a$,  never-ending oscillations have been detected \cite{AS17,AS18} in the region $\frac{1}{2} < a \leq 1$ and $ 0<\lambda\leq \lambda_c(a)$.


In this Letter, we consider a slightly simpler class of processes, and provide much stronger evidence for never-ending oscillations. We consider systems in which each aggregation event involves at least one monomer:
\begin{equation}
\label{A-def}
[1] \oplus [s]~ \xrightarrow{\text{$A_s$}} ~[1+s]
\end{equation}
This naturally occurs if only monomers are mobile as it happens, e.g., in monolayer growth \cite{Zan,Villain}. The shattering is assumed to be spontaneous 
\begin{equation}
\label{break}
[s]~ \xrightarrow{\text{$B_s$}} ~\underbrace{[1]+\cdots+[1]}_{s}
\end{equation}
rather than the collision-induced shattering \eqref{S-def}. The governing equations read 
\begin{subequations}
\begin{align}
\label{ns}
\frac{dc_s}{dt} &=c_1[A_{s-1} c_{s-1}-A_s c_s]-B_s c_s, \quad s\geq 2\\
\label{n1}
\frac{dc_1}{dt} &=\sum_{s\geq 2}^\infty s B_s c_s
-2A_1 c_1^2 -c_1\sum_{s\geq 2} A_s c_s
\end{align} 
\end{subequations}
The system is closed, so the mass density is conserved:
\begin{equation}
\label{mass}
M = \sum_{s=1}^\infty sc_s(t) \equiv \text{const}
\end{equation}

There is no natural relation between spontaneous shattering rates $B_s$ and collision-controlled addition rates $A_s$. Pure addition processes with rates $A_s = s^a$ have been investigated \cite{bk}. The merging rate cannot grow faster than mass, so on the physical grounds $a\leq 1$. Furthermore, addition processes with rates  $A_s = s^a$  and $a >1$ are ill-defined due to instantaneous gelation \cite{bk,instant}. Thus it is reasonable to choose $a\leq 1$. Most interesting behaviors are anticipated near the maximal growth exponent $a=1$. Hence we take $A_s=s$ and recast \eqref{ns}--\eqref{n1} into
\begin{subequations}
\begin{align}
\label{ns-1}
\frac{dc_s}{dt} &=c_1[(s-1) c_{s-1}- s c_s]-B_s c_s, \quad s\geq 2\\
\label{n1-1}
\frac{dc_1}{dt} &=\sum_{s\geq 2}^\infty s B_s c_s
- c_1^2 - c_1
\end{align} 
\end{subequations}
where we additionally set $M=1$. These equations are too general, so we further specialize Eqs.~\eqref{ns-1}--\eqref{n1-1} to a class of algebraic break-up rates
\begin{equation}
\label{Bs-beta}
B_s=B s^\beta
\end{equation}


Suppose that the system  reaches a steady state.  From \eqref{ns-1} we find that the stationary size
distribution obeys  $c_s=c_{s-1}(s-1)/(s+B_s/n_1)$, from which
\begin{equation}
\label{c-SS}
\frac{c_s}{c_1} =\prod_{j=2}^s \frac{j-1}{j+B_j/c_1}
\end{equation}  
This is valid for arbitrary break-up rates $B_s$, modulo of course the assumption that a steady state is reached. 

Using \eqref{c-SS} with $\beta<0$, we deduce $c_s\sim s^{-1}$. The tail must decay faster than $s^{-2}$ to agree with mass conservation, $\sum_{s\geq 1}sc_s=1$. Thus the assumption that the system reaches the steady state is erroneous when $\beta<0$. Instead, the typical size grows indefinitely, i.e., shattering rates with $\beta<0$ are too weak to counter-balance growth via addition. Similar coarsening behaviors have been observed in a few other aggregation-fragmentation processes, see e.g. \cite{Colm-PK}. Leaving the complete analysis of the behavior in the $\beta<0$ for future, let us consider the behavior when $|\beta|\gg 1$ limit. The shattering rates vanish when $\beta=-\infty$, so in the first stage, we drop them. In this situation, the monomers quickly disappear. If $c_s(0)=\delta_{s,1}$, one gets \cite{bk}
\begin{equation}
\label{nst}
c_s(t)=\frac{(1-e^{-t})^{s-1}-s^{-1}(1-e^{-t})^s}{(2-e^{-t})^s}
\end{equation}  
from which $c_s(\infty)=\left(1-s^{-1}\right)\cdot 2^{-s}$. Thus without shattering, the system freezes into a stationary state with no monomers and an exponential cutoff in the size distribution. In the second stage, dimers start to break, while the heavier clusters remain stable. The time scale for this second stage is $O(2^{-\beta})$. At the end of the second stage, there are no monomers and dimers. In the third stage, trimers start to break. The corresponding time scale is $O(3^{-\beta})$. At the end of this stage, there are no monomers, dimers, and trimers.  This continues demonstrating coarsening in the $\beta\to -\infty$ limit.

\begin{figure}[ht]
\begin{center}
\includegraphics[width=0.35\textwidth]{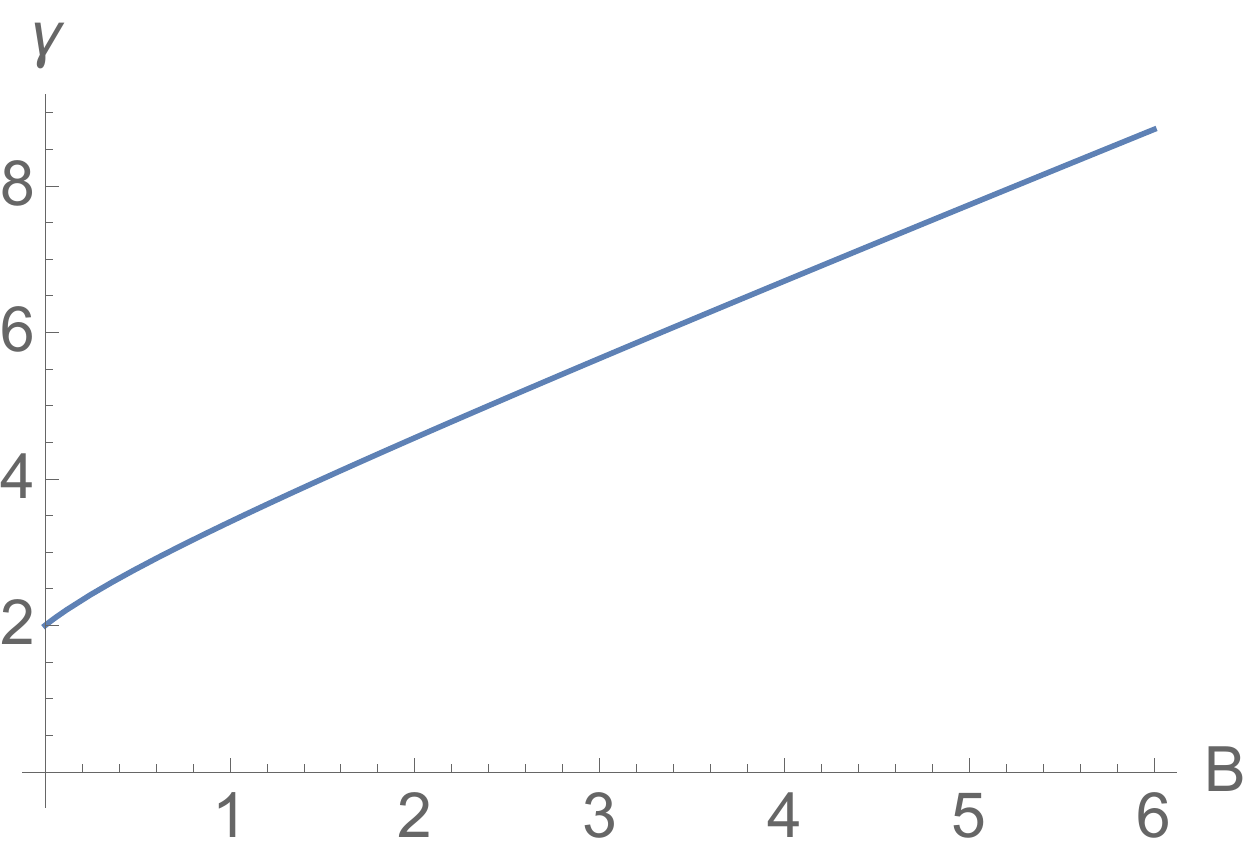}
\caption{The plot of the exponent $\gamma$ given by \eqref{def:gamma}.}
\label{fig:gamma}
 \end{center}
\end{figure}

Therefore oscillations may occur only when $\beta\geq 0$, so in the following we focus on this range.  From \eqref{c-SS} we deduce the asymptotic behaviors 
\begin{equation}
\label{c-SS-b}
\frac{c_s}{c_1} \propto 
\begin{cases}
(c_1/B)^s (s!)^{-(\beta-1)}                & \beta>1 \\
s^{-1}\exp[-s^\beta B/\beta c_1]     & 0<\beta<1\\
s^{-1-B/c_1}                                    & \beta=0
\end{cases}
\end{equation}  
Qualitative changes occur at $\beta=1$ and $\beta=0$. At these marginal cases one can obtain more precise results: 
\begin{equation}
\label{c-SS-1}
\frac{c_s}{c_1} = s^{-1}(1+B/c_1)^{1-s}, \quad c_1 =\frac{\sqrt{B^2+4B}-B}{2}
\end{equation}
when $B_s = B s$, while when $B_s = B$ we get 
\begin{equation}
\label{c-SS-0}
\frac{c_s}{c_1} = \frac{\Gamma(s) \Gamma(2+B/c_1)}{\Gamma(s+1+B/c_1)}
\end{equation}
with
\begin{equation}
\label{mon-0}
c_1= \frac{b-1-B}{2}\,, \quad b\equiv \sqrt{B^2+6B+1}
\end{equation}
The steady state \eqref{c-SS-0} has an algebraic tail, $c_s \sim s^{-\gamma}$ for $s\gg 1$, with
\begin{equation}
\label{def:gamma}
\gamma=1+\frac{B}{c_1} = \frac{b-1+B}{b-1-B}
\end{equation}
The exponent $\gamma$ is an increasing function of the amplitude $B$. Starting from $\gamma = 2$ for $B = 0$ it grows asymptotically  as $B + 1$ for $B \gg 1$; see Fig.~\ref{fig:gamma}. 

The $\beta=0$ case is special since Eq.~\eqref{n1-1} is particularly simple in this situation:
\begin{equation}
\label{n1-10}
\frac{dc_1}{dt} =B(1-c_1)- c_1^2 -c_1
\end{equation}
From this closed equation, we deduce that \eqref{mon-0} is indeed a stable fixed point for the monomer density. If $c_1(0)=1$, the explicit expression for the monomer density reads 
\begin{equation}
\label{mon-0-t}
c_1(t)= c_1+\frac{b (1-c_1)^2 }{2\,e^{b t}-(1-c_1)^2}
\end{equation}
with $c_1\equiv c_1(\infty)$ and $b$ given by \eqref{mon-0}. The remaining equations \eqref{ns-1} can be re-written as
\begin{equation}
\label{ns-10}
\frac{dn_s}{d\tau} =(s-1) n_{s-1}- s n_s, \quad s\geq 2
\end{equation}
with $n_s(\tau)=e^{Bt}\,c_s(t)$ and $\tau=\int_0^t dt'\,c_1(t')$. These equations with already known $n_1(\tau)=e^{Bt}\,c_1(t)$, where $c_1(t)$ is given by \eqref{mon-0-t}, can be solved recurrently from which one verifies the stability of the fixed point \eqref{c-SS-0}--\eqref{mon-0}. 


\begin{figure}[ht]
\begin{center}
\includegraphics[width=0.4\textwidth]{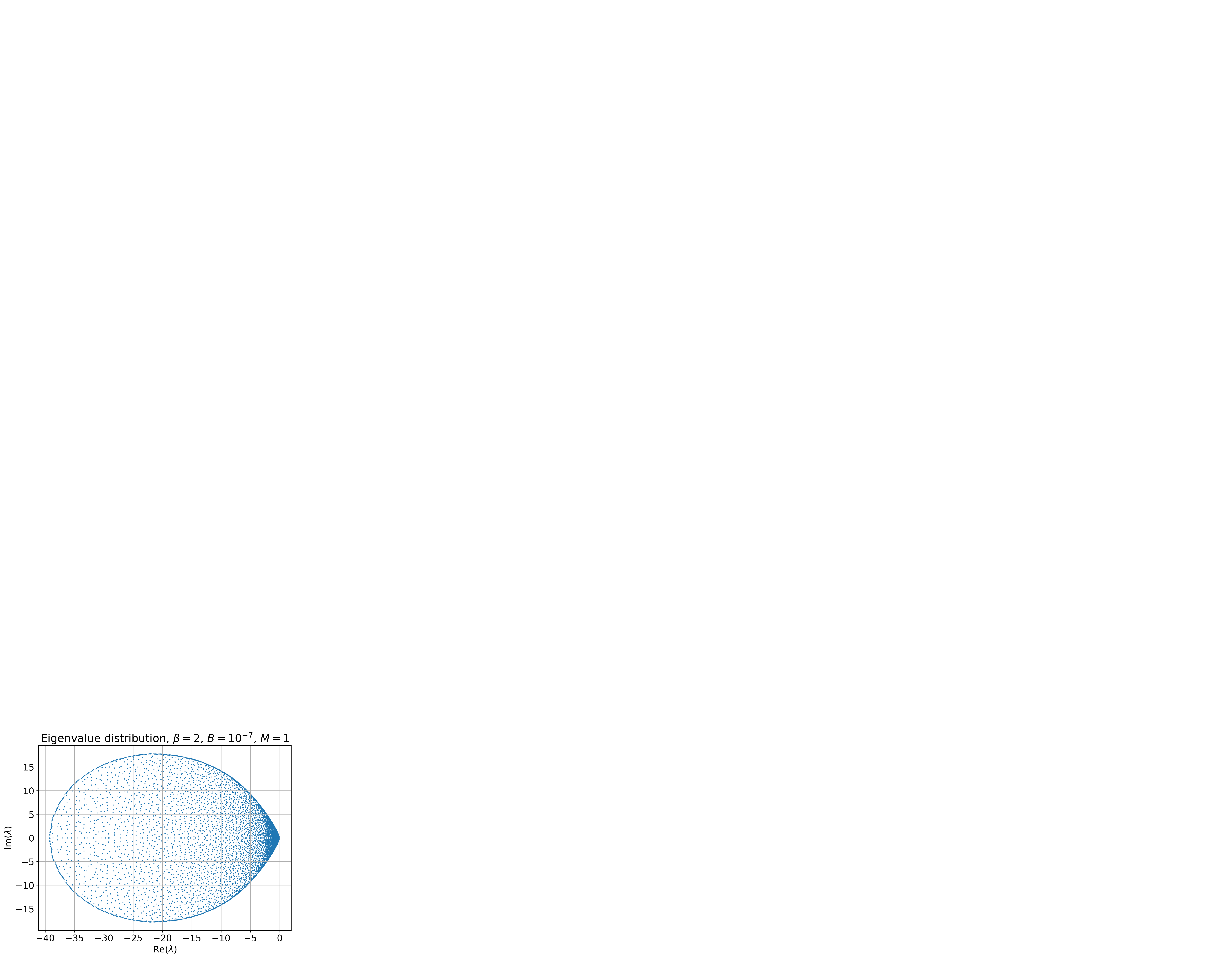}
\caption{The eigenvalues of the linearized system truncated to $N = 5000$ equations. The eigenvalues have negative real parts, and concentrate close to zero, with a possible exceptional pair of eigenvalues that has a positive real part. This pair appears in region $\mathcal{U}$ of the parameter space, and it is responsible for  oscillations.  }
\label{fig:spectrum}
\end{center}
\end{figure}

\begin{figure}[ht]
\begin{center}
\includegraphics[width=0.4\textwidth]{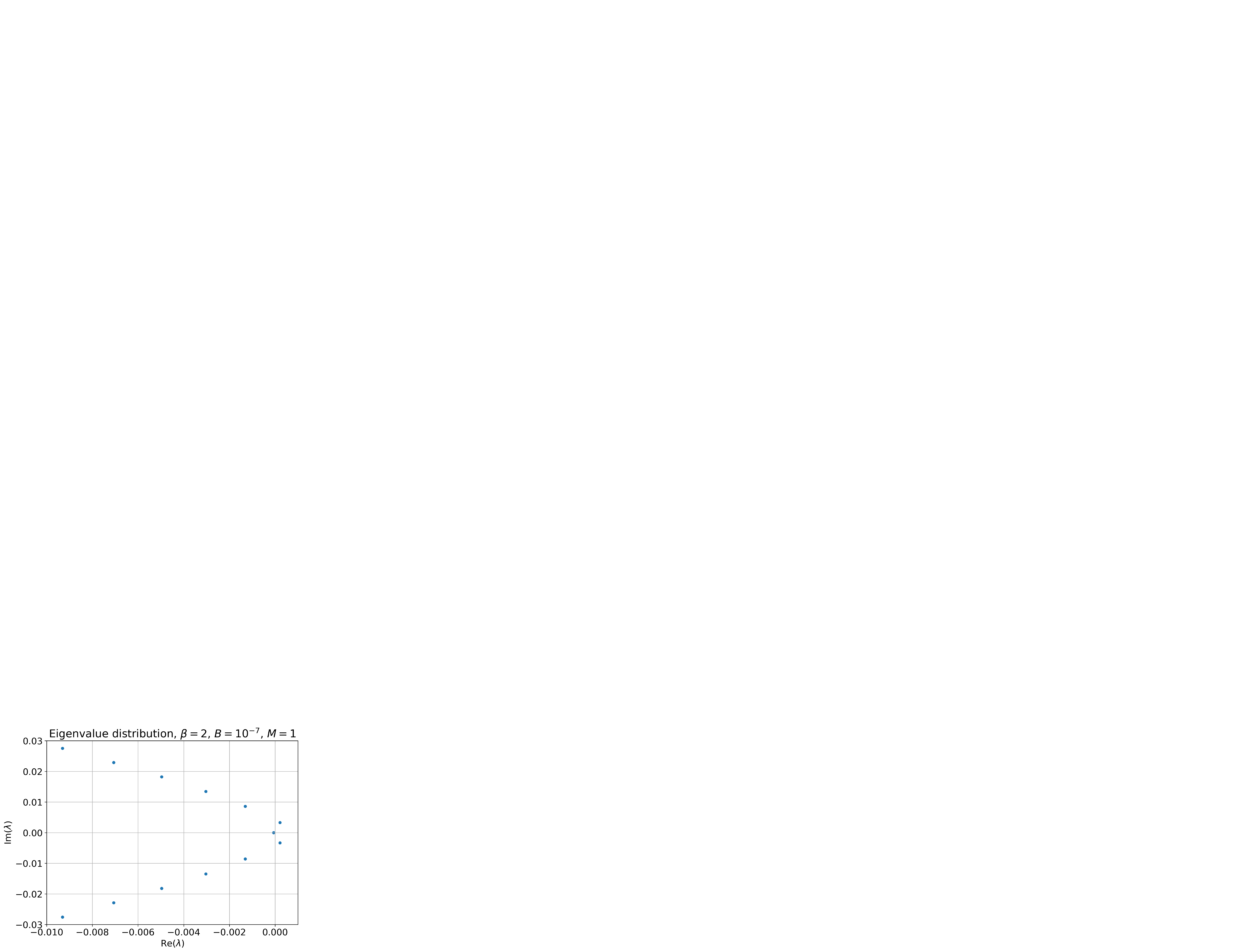}
\caption{The eigenvalues near $\lambda=0$, the same parameters as in Fig.~\ref{fig:spectrum}. The pair of eigenvalues with positive real part that causes oscillations is clearly visible.  }
\label{fig:spectrum_zoom}
\end{center}
\end{figure}

\begin{figure}[ht]
\begin{center}
\includegraphics[width=0.4\textwidth]{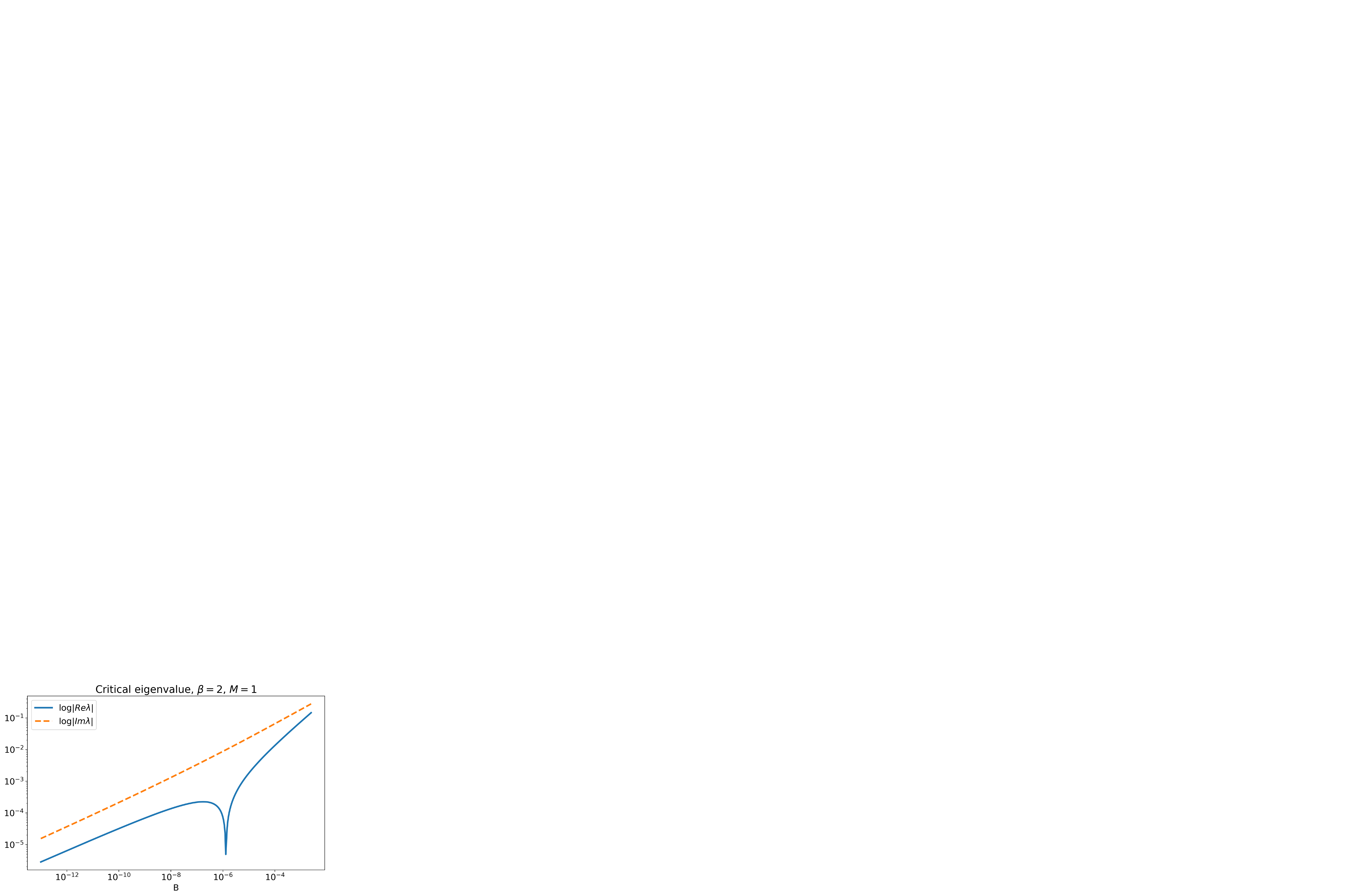}
\caption{Real and imaginary parts of the eigenvalue that crosses the imaginary axis as $B$ varies. The cusp in the plot of $\log |\text{Re} \lambda|$ corresponds to Hopf bifurcation: The critical eigenvalue changes its sign, and a limit cycle is born.}
\label{fig:eigval}
\end{center}
\end{figure}

\begin{figure}[t]
\begin{center}
\includegraphics[width=0.4\textwidth]{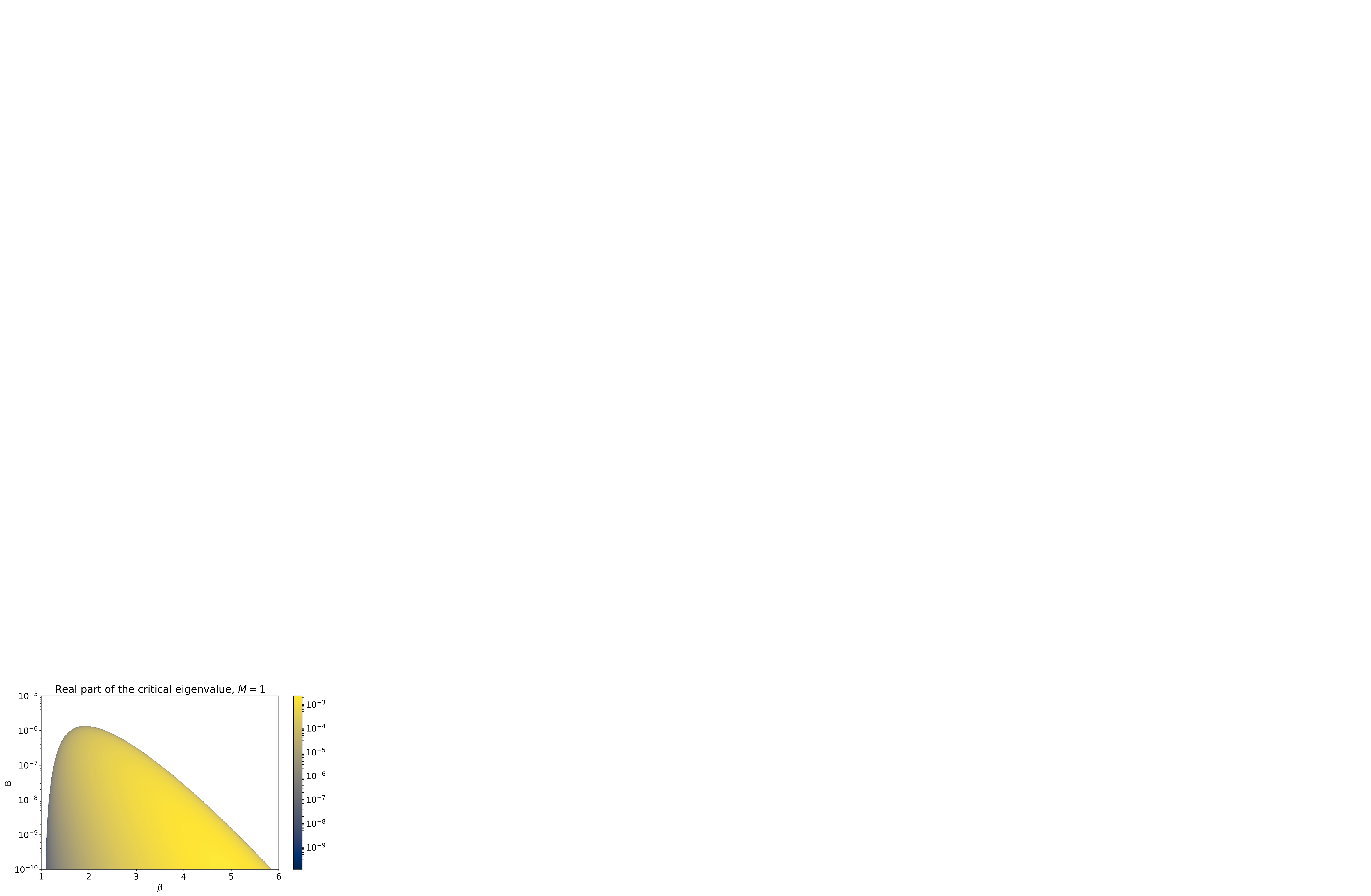}
\caption{The region of the $(\beta, B)$ plane where unstable eigenvalues exist. It means that these values of the parameters correspond to the birth of oscillations.}
\label{fig:region}
\end{center}
\end{figure}

The stability of the steady state is difficult for theoretical analysis when $\beta > 0$. We study it by perturbing Eqs. \eqref{ns-1}--\eqref{n1-1} near the fixed point in a way that preserves the mass density, and we explore the eigenvalues of the corresponding linearized aggregation operator. After truncating the infinite system to $N$ equations, we find that the eigenvalues of the corresponding Jacobian matrix have negative real parts, with at most one exceptional pair of eigenvalues with a positive real part (see Fig.~\ref{fig:spectrum}--\ref{fig:spectrum_zoom}). This pair is present in a certain region in the parameter space
\begin{equation}
\label{tongue}
\mathcal{U} = \{(\beta, B)|\, \beta > 1, ~ 0 < B < B_{\text{crit}}(\beta) \}
\end{equation}
This occurs only for sufficiently large $N$. The steady state loses stability via Hopf bifurcation when $B$ crosses the critical value $B_{\text{crit}}(\beta)$ and enters region $\mathcal{U}$. This leads to the birth of a stable limit cycle. The imaginary part of the critical eigenvalue decreases monotonically, ignorant to the bifurcation, as $B$ decreases (Fig.~\ref{fig:eigval}). The real part changes its behavior once the eigenvalue becomes unstable: $\text{Re}(\lambda)$ keeps growing for a short while before reaching its maximum value and then decays monotonically. Figure \ref{fig:region} shows the transition curve $B_{\text{crit}}(\beta)$ in the parameter space. In particular, it shows that there is a singularity at $\beta = 1$, whose existence is connected with the qualitative changes in the steady state \eqref{c-SS-b}--\eqref{c-SS-1}. In our numerical experiments we exploit the structure of the Jacobian and use the inverse power method \cite{tee-brief-book} to find unstable eigenvalues (see SM for details).

\begin{figure}[t]
\begin{center}
\includegraphics[width=0.45\textwidth]{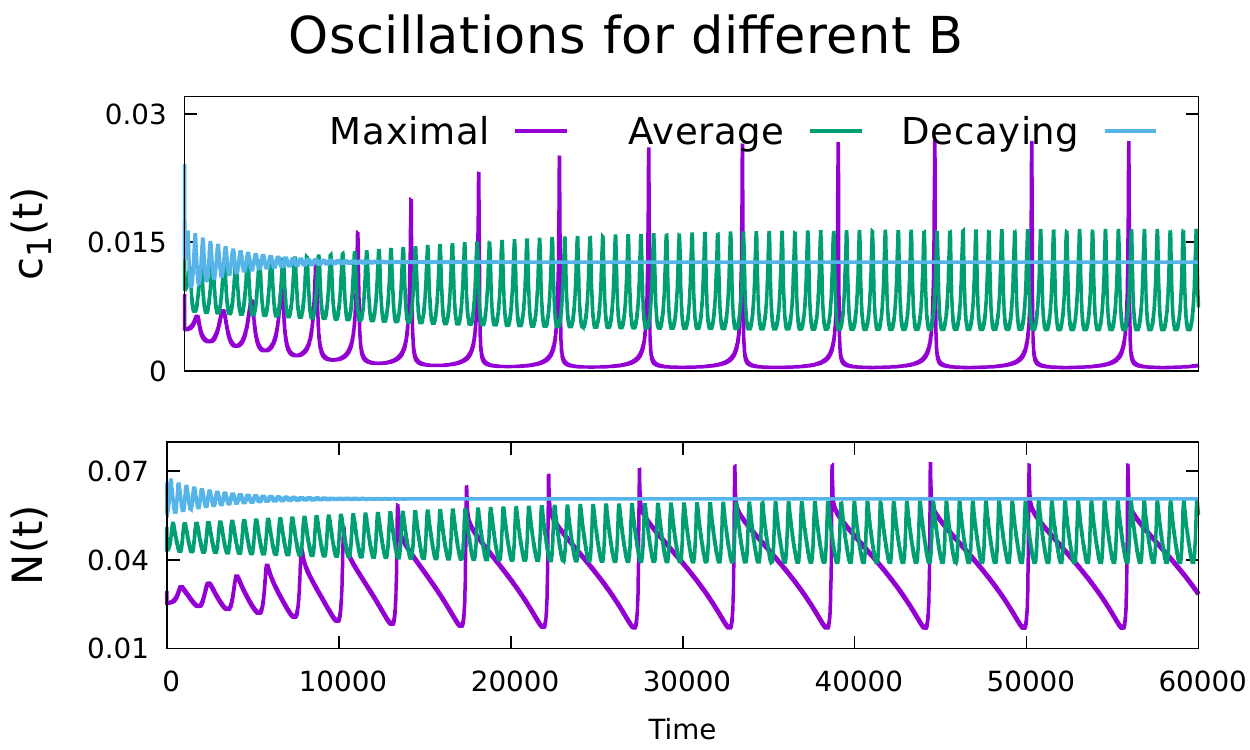}
\caption{Oscillatory regimes for monomers $c_1(t)$ and total density $N(t) = \sum_{k=0}^{\infty} c_k$ for different values of $B$ with $\beta = 2$ and unit mass density. The oscillations decay when $B = 3.1622776602 \cdot 10^{-6}$ [Decay], persist when $B = 1.2195704602 \cdot 10^{-6}$ [Average], and have the largest possible (for $\beta = 2$ and unit mass) amplitude when $B = 1.668100537 \cdot 10^{-7}$ [Maximal]. In all three cases we used initial conditions as in \eqref{ic}.}
\label{fig:oscillations}
\end{center}
\end{figure}

We carried out numerical simulations to study the oscillatory solutions of the system truncated to $N$ equations, with $N$ sufficiently large to ensure that mass conservation is held on each iteration with machine precision. This makes the finite system numerically indistinguishable from the infinite one. The results are presented in Fig. \ref{fig:oscillations} for multiple values of $B$ with $\beta = 2$ fixed. The initial condition was taken as a perturbation of the steady state \eqref{c-SS}--\eqref{c-SS-b} that preserves its mass density:
\begin{equation}
\label{ic}
    \tilde{c}_1 = c_1 + 1.8 c_2, \quad \tilde{c}_2 = 0.1 c_2, \quad \tilde{c}_s = c_s.
\end{equation}
Comparing Figs. \ref{fig:oscillations} and \ref{fig:eigval} we see that the oscillations die out when $B$ is above the critical value $B_{\text{crit}}(2)$ and persist when $B$ is below it. As $B$ continues to decrease, the amplitude of the oscillations at first grows, reaches its maximum, and starts decaying to zero. The frequency of the oscillations decreases monotonically with $B$.

How many equations should one take to see oscillations? In our numerical experiments, we have observed that Eqs. \eqref{ns-1}--\eqref{n1-1} cease to have unstable eigenvalues for parameters from the region \eqref{tongue} when $N$ is not big enough. The ``big enough" grows as $B$ tends to zero or $\beta$ tends to one, together with the effective length of the stationary distribution. At another extreme, one can consider Eqs.~\eqref{ns}--\eqref{n1} and set $A_s$ to zero for $s \geq 3$. Finding limit cycles is difficult even for such simple systems of two coupled ODEs. Several tools allow one to rule out the limit cycles or prove their existence \cite{Perko,Chris,Strogatz}. In our case, the application of the Dulac criterion shows the absence of limit cycles independently of the rates (see SM). Recent results on Hopf bifurcation in a finite Becker-D\"oring exchange model also show that the number of ODEs in such finite systems has to be sufficiently large to obtain oscillatory solutions \cite{Pego20}. This perhaps explains why despite years of searching, the oscillatory solutions have not been observed.


To summarize, we have found oscillatory solutions in the realm of addition-shattering models \eqref{ns-1}--\eqref{n1-1} with algebraic break-up rates \eqref{Bs-beta}. These solutions are born through the Hopf bifurcation mechanism: The steady states exist whenever $\beta \geq 0$, but become unstable for parameters from \eqref{tongue} and give birth to never-ending oscillations via Hopf bifurcation. Oscillatory solutions in other models have been detected recently \cite{AS17,AS18,Pego20}. For instance, Hopf bifurcation has been found in a finite Becker-D{\"o}ring system with constant kinetic coefficients \cite{Pego20}. Our infinite system with algebraically growing rates also exhibits oscillatory solutions, at least the numerical evidence is very convincing.

In a class of addition-shattering processes that we investigated, persistent oscillations occur in a small region of the phase space; the same holds for the model studied in \cite{AS17,AS18}. This rarity is similar to the empirical evidence with limit cycles in planar systems with quadratic polynomials. Systems with up to four limit cycles are known  \cite{H4_1,H4_2,Kuz1,Kuz}. If $H(2)=\infty$, there exist planar systems with quadratic polynomials and an arbitrary number of limit cycles. The rule of thumb, however,  is that a ``generic" planar system has no limit cycles (see \cite{Kuz}). The same seemingly holds for aggregation-fragmentation systems. Limit cycles are very rare, and systems with more than one limit cycle are currently unknown. Another avenue for future work is to seek oscillations in systems with standard binary fragmentation. Among the biggest challenges is providing rigorous proof of persistent oscillations in an infinite system and finding chaos.

\bigskip\noindent
The work of S.S.B and S.A.M was partly supported by Moscow Center for Fundamental and Applied Mathematics (agreement No. 075-15-2019-1624).

\bibliographystyle{apsrev}

\begin{thebibliography}{99}
\bibitem{LC:def}
       By definition, the limit cycle is an isolated closed orbit of a vector field. For linear vector fields, a continuous family 
       of closed orbits is possible, but these orbits are not isolated; e.g., for the harmonic oscillator described
       by \eqref{PQ} with $P=y$ and $Q=-x$, all circles centered at the origin are orbits.
\end{thebibliography}

\newpage
\clearpage

\pagenumbering{gobble}
\def\beq{\begin{eqnarray}}
\def\eeq{\end{eqnarray}}
\def\be{\begin{equation}}
\def\ee{\end{equation}}
\def\eq{&=&}
\def\ct{\cite}

\def\bm{\begin{math}}
\def\me{\end{math}}
\def\bi{\bibitem}
\def\vr{\vec r}
\def\lt{L(t)}
\def\al{a(L,T)}
\def\hf{\frac{1}{2}}
\def\ap{\alpha}
\def\rb{\right )}
\def\la{\langle}
\def\ra{\rangle}
\def\lbr{\left [}
\def\rbr{\right ]}
\def\del{\partial}
\def\grad{\nabla}
\def\ul{\underline}
\def\etal{{\it et al.}}
\def\lra{\leftrightarrow}
\def\rar{\rightarrow}
\def\lb{\label}
\def\q{\quad}
\def\qq{\qquad}
\newcommand \nn {\nonumber}
\newcommand \bei {\begin{itemize}}
\newcommand \eei  {\end{itemize}}
\newcommand \ii    {\item}
\newcommand \nt   {\nonumber \\ }
\def\Real{\mathbb{R}}
\def\Clx{\mathbb{C}}

\section*{Supplementary material: Hopf bifurcation in addition-shattering kinetics}

\subsection{Truncated Models and Dulac function}
\label{sec:truncate}

In aggregation-fragmentation processes \eqref{AF:gen}, never-ending oscillations have not been found analytically. This is not surprising as it requires a solution of an infinite set of non-linear ODEs. To appreciate the existence of oscillations one can first seek such solutions in truncated models in which the matrix elements $K_{ij}$ vanish when $i+j$ is sufficiently large. 

We define $m-$truncated models by requiring
\begin{equation}
K_{ij}=0 \quad\text{when}\quad i+j>m
\end{equation}
For such models, the system \eqref{AF:gen} of infinitely many ODEs reduces to ODEs for the densities $c_1,\ldots,c_m$. Taking into account mass conservation
\begin{equation}
\label{massSM}
\sum_{j=1}^m jc_j(t) = 1
\end{equation}
reduces the number of ODEs to $m-1$. Limit cycles are possible in a system of two (or more) ODEs. Thus in the truncated models, limit cycles 
may arise when $m\geq 3$. Chaos becomes (in principle) feasible when $m\geq 4$. 

The $m=3$ truncated model consists of three ODEs
\begin{subequations}
\begin{align}
&\frac{d c_1}{dt} = -2c_1^2 - 2Kc_1c_2 + 2F c_2 + G c_3
\label{1}\\
&\frac{d c_2}{dt} = c_1^2 - 2Kc_1c_2 - F c_2 + G c_3
\label{2}\\
&\frac{d c_3}{dt} = 2Kc_1c_2 -  G c_3
\label{3}
\end{align}
\end{subequations}
where we shortly write the matrices $||K_{ij}||$ and $||F_{ij}||$ with $i, j\leq 2$ as 
\begin{equation}
\label{KF:3}
||K_{ij}|| = 
\left( \begin{array}{cc}
2  & K \\
\\
K& 0
\end{array} \right),
\quad ||F_{ij}|| =
\left( \begin{array}{cc}
2F  & G \\
\\
G& 0 
\end{array} \right)
\end{equation}

Specializing \eqref{massSM} to $m=3$ we get $2c_2=1-c_1-3c_3$. Substituting this relation to \eqref{1} and \eqref{3} we obtain
\begin{subequations}
\begin{align}
\label{c1:eq}
\frac{d c_1}{dt} &= (F-\tfrac{1}{2}Kc_1)(1-c_1-3c_3) + G c_3   - 2c_1^2 \\
\frac{d c_3}{dt} &=   \tfrac{1}{2}Kc_1(1-c_1-3c_3) -  G c_3
\label{c3:eq}
\end{align}
\end{subequations}

Equations \eqref{c1:eq}--\eqref{c3:eq} are the most general equations for the truncated model with $m=3$. Indeed, \eqref{KF:3} are the most general rates  satisfying the symmetry requirement, we merely disregarded the pathological case $K_{11}=0$ and set $K_{11}=2$ by rescaling the time variable. 

On the physical grounds, the rates are positive. Hence the parameters should  lie inside the octant 
\begin{equation}
\label{space}
\mathbb{R}^3_+=\{(F,G,K)|\,F > 0, ~ G > 0,  ~K > 0\}
\end{equation}
We are also interested in the behavior inside the triangle
\begin{equation}
\label{triangle}
\mathcal{T} = \{(c_1, c_3)|\, c_1\geq 0, ~ c_3\geq 0, ~ c_1+3c_3\leq 1\}
\end{equation}
Indeed, the densities are non-negative and mass conservation written as $2c_2=1-c_1-3c_3\geq 0$ explains the last inequality. If the system starts inside the triangle, it forever remains there.

We now rule out the existence of limit cycles for the system \eqref{c1:eq}--\eqref{c3:eq} using the Dulac criterion \cite{Perko,Chris,Strogatz}. For the general planar system \eqref{PQ}, the  Dulac criterion asserts that if there exists a smooth function $D(x,y)$ in a simply-connected domain $\mathcal{D}\subset \mathbb{R}^2$ such that the Dulac function 
\begin{equation}
\mathbb{D}\equiv  \partial_x[DP]+\partial_y[DQ]
\end{equation}
has the same sign throughout $\mathcal{D}$, there are no closed orbits lying entirely in $\mathcal{D}$. Choosing $D(c_1,c_3)=1$ and $\mathcal{T}$ as the domain, one computes the Dulac function 
\begin{equation}
\mathbb{D} =-F-G -K(c_1+c_2)-4c_1
\end{equation}
Thus $\mathbb{D} < 0$ assuring the absence of limit cycles for the general truncated system \eqref{c1:eq}--\eqref{c3:eq}. 

Similarly, a limit cycle is impossible for the arbitrary addition-shattering process truncated to $m=3$. Indeed, Eqs.~\eqref{ns}--\eqref{n1} turn into a planar system 
\begin{equation}
\label{c12-eq}
\dot c_1 = P, \qquad \dot c_2 = Q
\end{equation}
with quadratic polynomials 
\begin{equation}
\label{R12}
\begin{split}
P &= (B_2 - \tfrac{1}{2}A_2 c_1)(1-c_1-3c_3)+3B_3 c_3 -2A_1c_1^2\\
Q &=\tfrac{1}{2}A_2 c_1(1-c_1-3c_3)-B_3 c_3
\end{split}
\end{equation}
depending on four positive rates: $A_1,A_2,B_2,B_3$.  Choosing again $D(c_1,c_3)=1$ and the triangle $\mathcal{T}$ as the domain, we compute the corresponding Dulac function 
\begin{equation}
\frac{\partial P}{\partial c_1}+ \frac{\partial Q}{\partial c_2}=-B_2-B_3-A_2(c_1+c_2)-4A_1 c_1
\end{equation}
and find that it is negative assuring the absence of limit cycles for the truncated system \eqref{c12-eq}--\eqref{R12}.

\subsection{Linearizing Eqs. \eqref{ns-1}--\eqref{n1-1} about the steady state}

Equation \eqref{c-SS} asserts that the stationary size distribution is uniquely determined by the density $c_1$ of monomers. The mass density
\begin{equation}
\label{ss-mass}
M = \sum_{s \geq 1} s c_s = c_1 \sum_{s \geq 1} s \prod_{j = 2}^{s} \frac{j - 1}{j + B_j / c_1}
\end{equation}
increases monotonically with $c_1$ and thus for every value of $M$, the system \eqref{ns-1}--\eqref{n1-1} has at most one steady state. To numerically find the steady state with a given mass density it suffices to solve a nonlinear equation \eqref{ss-mass}.

Owing to mass conservation, the sets of equal-mass size distributions are invariant for \eqref{ns-1}--\eqref{n1-1}. And when we talk about the birth of limit cycles we always confine the system to distributions of fixed mass $M$, which we can choose to be unity since the system remains unchanged under scaling
\begin{equation}
\label{scaling}
c_s \mapsto M c_s, \quad B_s \mapsto M B_s, \quad t \mapsto \frac{1}{M} t
\end{equation}
To preserve the total mass, we consider the following perturbations $\{ x_s \}$ of the steady state:
\begin{equation}
\label{perturb-mass}
\sum_{s \geq 1} s x_s(t) = 0.
\end{equation}
In the vicinity of the steady state, equations \eqref{ns-1}--\eqref{n1-1} read
\begin{subequations}
\begin{multline}
\label{xs-1}
\frac{dx_s}{dt} = (c_1 + x_1)[(s-1) x_{s-1}- s x_s] -\\ B_s\left[\frac{c_s}{c_1}x_1 - x_s\right], \quad s\geq 2
\end{multline}
\begin{equation}
\label{x1-1}
\frac{dx_1}{dt} = -(c_1 + x_1)x_1 - \sum_{s \geq 2} s B_s \left[\frac{c_s}{c_1}x_1 - x_s \right]
\end{equation}
\end{subequations}
Dropping nonlinear terms in Eqs.~\eqref{xs-1}--\eqref{x1-1} we arrive at
\begin{subequations}
\begin{multline}
\label{xs-1-lin}
\frac{dx_s}{dt} = c_1[(s-1) x_{s-1}- s x_s] -\\ B_s\left[\frac{c_s}{c_1}x_1 - x_s\right], \quad s\geq 2
\end{multline}
\begin{equation}
\label{x1-1-lin}
\frac{dx_1}{dt} = -c_1x_1 - \sum_{s \geq 2} s B_s \left[\frac{c_s}{c_1}x_1 - x_s \right]
\end{equation}
\end{subequations}

\subsection{Numerical approach for evaluation of the spectrum}

Fix $N$ and consider the first $N$ equations of \eqref{xs-1}--\eqref{x1-1}. Such truncation obviously breaks the mass conservation law but it holds with machine precision provided $N$ is sufficiently large, making the finite system numerically indistinguishable from the infinite one. 

The elements of the Jacobian matrix $\mathbf{J} \in \Real^{N \times N}$ are given by
\begin{equation}
\label{jac}
\mathbf{J}(i, j) = 
\begin{cases}
    -c_1 - \sum_{s \geq 2} s B_s \frac{c_s}{c_1}, & i = 1,\,j = 1 \\    
    j B_j, & i = 1,\,j > 1 \\
    c_1 - B_2 \frac{c_2}{c_1}, & i = 2,\,j = 1 \\
    -B_i \frac{c_i}{c_1}, & i > 2,\,j = 1 \\
    (i - 1) c_1, & i > 2,\,j = i - 1 \\ 
    B_i - i c_1, & i \geq 2,\,j = i \\
    0, & \text{otherwise}
\end{cases}
\end{equation}
For moderate values of $N$ we can compute the complete spectrum $\sigma(\textbf{J})$ of $\mathbf{J}$ with the help of standard LAPACK procedures (or their wrappers as in \texttt{numpy}). For example, Fig. \ref{fig:spectrum} of the main text was obtained this way for $N = 5000$.

However, with Hopf bifurcation in mind, we are not interested in the whole spectrum of $\mathbf{J}$ but only in its eigenvalues that invade the complex half-plane with a positive real part, $\text{Re} \lambda > 0$. When $\beta$ is close to unity, $N$ gets as big as $10^7$ making the computation of all the eigenvalues not only superfluous but highly inefficient. 

Instead, we can use the so-called \textit{inverse iterations} (or \textit{inverse power method}) that allow one to find the eigenvalue closest to a given complex number $\mu \in \Clx$ and its corresponding eigenvector. The iterations start from an initial vector $\textbf{v}_0 \in \Clx^N$ that is typically chosen to be random unless some a priori information is available. At each iteration, the algorithm solves a linear system of equations and normalizes a vector:
\begin{equation}\label{inv_iter}
    \textbf{u}_k = (\textbf{J} - \mu \textbf{I}_N)^{-1} \textbf{v}_{k-1}, \quad \textbf{v}_k = \frac{\textbf{u}_k}{\| \textbf{u}_k \|}
\end{equation}
The resulting vector is an approximate eigenvector of $\textbf{J}$:
\[
    \textbf{J} \textbf{v}_k \approx \lambda_k \textbf{v}_k, \quad \lambda_k \approx \text{argmin}_{\lambda \in \sigma(\textbf{J})} |\lambda - \mu|
\]
This method converges fast and very few iterations are needed when $\mu$ is close to the desired eigenvalue.

\begin{figure}[h]
\begin{center}
\includegraphics[width=0.45\textwidth]{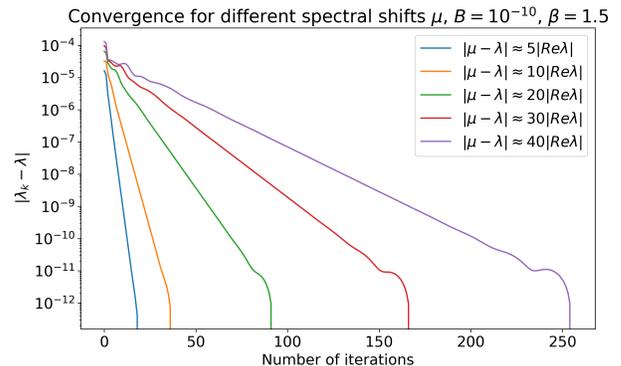}
\caption{The inverse power method converges geometrically and its rate of convergence depends on how close the spectral shift $\mu$ is to the eigenvalue $\lambda$ we are computing. The error is measured as $|\lambda - \lambda_k|$, where $\lambda_k$ is the result of the $k$-th iteration. We used $B = 10^{-10}$, $\beta = 1.5$, and $N = 10^7$.}
\label{fig:inv_pwr}
\end{center}
\end{figure}

The computational complexity of the algorithm stems from the need to solve a linear system of equations at each iteration \eqref{inv_iter}. To make the iterations efficient, we exploit the structure of the Jacobian \eqref{jac}. Matrix $\mathbf{J}$ is extremely sparse and has the following template:
\begin{equation}\label{jac_template}
    \mathbf{J} =
    \begin{bmatrix}
        \times & \times & \times & \times & \ldots & \times & \times & \times \\
        \times & \times &   0    &   0    & \ldots &   0    &   0    &   0    \\
        \times & \times & \times &   0    & \ldots &   0    &   0    &   0    \\
        \times &   0    & \times & \times & \ldots &   0    &   0    &   0    \\
        \vdots & \vdots & \vdots & \vdots & \ddots & \vdots & \vdots & \vdots \\
        \times &   0    &   0    &   0    & \ldots & \times & \times &   0    \\
        \times &   0    &   0    &   0    & \ldots &   0    & \times & \times \\
    \end{bmatrix},
\end{equation}
where $\times$ denotes nonzero elements. Matrices of this form \eqref{jac_template} admit an exceptionally pleasant upper-lower triangular factorization $\mathbf{J} = \mathbf{U} \mathbf{L}$ with
\begin{subequations}
\begin{equation}\label{jac_u}
    \mathbf{U} =
    \begin{bmatrix}
        \times & \times & \times & \times & \ldots & \times & \times & \times \\
          0    & \times &   0    &   0    & \ldots &   0    &   0    &   0    \\
          0    &   0    & \times &   0    & \ldots &   0    &   0    &   0    \\
          0    &   0    &   0    & \times & \ldots &   0    &   0    &   0    \\
        \vdots & \vdots & \vdots & \vdots & \ddots & \vdots & \vdots & \vdots \\
          0    &   0    &   0    &   0    & \ldots &   0    & \times &   0    \\
          0    &   0    &   0    &   0    & \ldots &   0    &   0    & \times \\
    \end{bmatrix}
\end{equation}
and
\begin{equation}\label{jac_l}
    \mathbf{L} =
    \begin{bmatrix}
        \times &   0    &   0    &   0    & \ldots &   0    &   0    &   0    \\
        \times & \times &   0    &   0    & \ldots &   0    &   0    &   0    \\
        \times & \times & \times &   0    & \ldots &   0    &   0    &   0    \\
        \times &   0    & \times & \times & \ldots &   0    &   0    &   0    \\
        \vdots & \vdots & \vdots & \vdots & \ddots & \vdots & \vdots & \vdots \\
        \times &   0    &   0    &   0    & \ldots & \times & \times &   0    \\
        \times &   0    &   0    &   0    & \ldots &   0    & \times & \times \\
    \end{bmatrix}
\end{equation}
\end{subequations}
This means that we can precompute the $\mathbf{U} \mathbf{L}$ factorization \eqref{jac_u}--\eqref{jac_l} of $\textbf{J} - \mu \textbf{I}_N$ and then solve two very sparse triangular systems per iteration \eqref{inv_iter}. 

We used this approach to compute the unstable region \eqref{tongue} in the parameter space as depicted in Fig. \ref{fig:region}. To further accelerate the computations, we employed parameter continuation: we took the approximate eigenvalue $\lambda$ and eigenvector $\textbf{v}$ corresponding to parameters $(\beta, B)$ as the spectral shift $\tilde{\mu}$ and starting vector $\tilde{\textbf{v}}_0$ for the adjacent parameters $(\tilde{\beta}, \tilde{B})$. This allowed us to process Jacobians of size $N = 10^7$ in reasonable time on a standard laptop.

Figure \ref{fig:inv_pwr} shows how the convergence of the inverse power method depends on the spectral shift $\mu$: The convergence is always geometrical, but its rate decreases when $\mu$ is far from the eigenvalue that we seek to compute. On a standard laptop, the computation takes 30-40 seconds in the worst case and less than 1 second in the best one. This phenomenon motivates one to exploit parameter continuation.

\subsection{The product kernel}

Aggregation-fragmentation processes in which both processes are collision-controlled, and each fragmentation event leads to complete shattering, have been studied in \cite{pnas2015,AS17,AS18} in the situation when the rates of aggregation and shattering events differ only by an amplitude:
\begin{equation}
\label{SK}
S_{ij} = \lambda K_{ij}
\end{equation}
This relation between the rates is natural since both aggregation and shattering are possible outcomes of the binary collision \cite{pnas2015}. The governing equations then read
\begin{subequations}
\begin{equation}
\label{AS:k}
\frac{dc_k}{dt} = \frac{1}{2}\sum_{i+j=k} K_{ij}\,c_i\,c_j-(1+\lambda)
c_k\sum_{j\geq 1} K_{kj}\,c_j 
\end{equation}
for $k\geq 2$, while the density of monomers satisfies 
\begin{eqnarray}
\label{AS:1}
\frac{dc_1}{dt} &=& - c_1\sum_{i\geq 1} K_{1,i}\,c_i + \lambda c_1\sum_{i\geq 2} i K_{1,i}\,c_i \nonumber\\
&+& \frac{ \lambda}{2}\sum_{i\geq 2} \sum_{j\geq 2} (i+j) K_{ij}\,c_i\,c_j
\end{eqnarray}
\end{subequations}
For the special class of rates
\begin{equation}
\label{K-a}
K_{ij} = (i/j)^a + (j/i)^a
\end{equation}
never-ending oscillations have been detected \cite{AS17,AS18} in the region
\begin{equation}
\{(a, \lambda)|\, \tfrac{1}{2} < a \leq 1, ~ 0<\lambda\leq \lambda_c(a)\}
\end{equation}
To appreciate the bounds $\frac{1}{2} < a \leq 1$ we note that aggregation equations with kernel  \eqref{K-a} and $a >1$ are ill-defined due to instantaneous gelation. Further, for aggregation equations with kernel  \eqref{K-a} driven by the constant input of monomers, the densities approach steady state values when $0\leq a<\frac{1}{2}$, while in the range $\frac{1}{2} < a \leq 1$ the densities evolve ad infinitum \cite{Colm-PK}. The shattering effectively acts as a source of monomers, and this qualitatively explains the appearance of $a_c = \frac{1}{2}$.

\begin{figure}[ht]
\begin{center}
\includegraphics[width=0.4\textwidth]{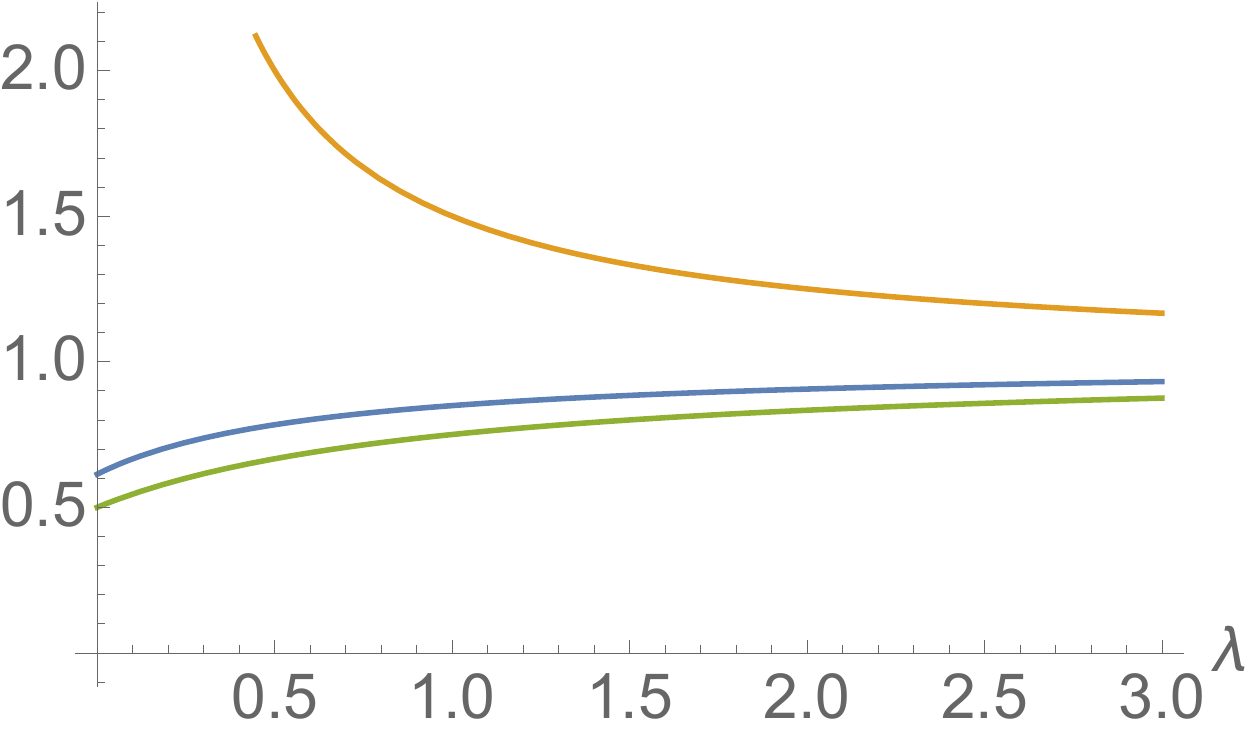}
\caption{Bottom to top: The stationary densities $c_1,~M_0,~M_2$. The density of monomers is 
$c_1=(1+2\lambda)/(2+2\lambda)$; the moments $M_0,~M_2$ are given by \eqref{MM}.}
\label{fig:cMM}
  \end{center}
\end{figure}

To gain insight into the behavior of the aggregation-shattering models satisfying \eqref{SK}, one may study kernels different from \eqref{K-a} and hopefully more amenable to analytical treatment. The product kernel
\begin{equation}
K_{ij} = ij
\end{equation}
is particularly well-known --- in the context of pure aggregation it provides the simplest description of gelation \cite{Flory,book}. For this kernel, Eqs.~\eqref{AS:k}--\eqref{AS:1} become 
\begin{subequations}
\begin{align}
\label{AS:k-product}
\frac{dc_k}{dt} &= \frac{1}{2}\sum_{i+j=k} ij c_i\,c_j-(1+\lambda)kc_k, \quad k\geq 2\\
\label{AS:1-product}
\frac{dc_1}{dt} &= -(1+\lambda)c_1+\lambda M_2
\end{align}
\end{subequations}
where  $M_2(t) =\sum_{j\geq 1} j^2 c_j(t)$ is the second moment, and the mass density is again set to unity: $\sum_{j\geq 1} j c_j(t) = 1$.

The system \eqref{AS:k-product}--\eqref{AS:1-product} does not admit solutions with never-ending oscillations. Instead, for every $\lambda>0$ solutions quickly approach to the steady state
\begin{equation}
\label{ck:ER}
c_k = \frac{1}{\sqrt{4\pi}}\,\frac{\Gamma\left(k-\frac{1}{2}\right)}{k\,\Gamma(k+1)}\,\frac{(1+2\lambda)^k}{(1+\lambda)^{2k-1}}
\end{equation}
The moments $M_p=\sum_{k\geq 1} k^pc_k$ approach to [see also Fig.~\ref{fig:cMM}]
\begin{equation}
\label{MM}
\begin{split}
M_0 & = 2+2(1+\lambda)\,\ln\frac{1+2\lambda}{2+2\lambda}\\
M_2 & = \frac{1+2\lambda}{2\lambda}\\
M_3 & = \frac{(1+2\lambda)(1+2\lambda+2\lambda^2)}{4\lambda^3}\\
M_4 & = \frac{(1+2\lambda)(3+12\lambda+18\lambda^2+12\lambda^3+4\lambda^4)}{8\lambda^5}
\end{split}
\end{equation}
etc. The tail of the distribution \eqref{ck:ER} is
\begin{equation}
c_k \sim k^{-5/2} e^{-\mu k}, \quad \mu =2\ln(1+\lambda)- \ln(1+2\lambda)
\end{equation}
Since $\mu\simeq \lambda^2$ as $\lambda\to +0$, the mass distribution decays algebraically, $c_k \sim k^{-5/2}$, when $1\ll k \ll \lambda^{-2}$. These analytical observations become extremely useful during the validation of accuracy of miscellaneous numerical methods.

\end{document}